\documentstyle[preprint,prl,aps,epsf]{revtex}

\begin{document}

\title{Low-Temperature Spin Dynamics of Doped Manganites: \\
roles of Mn-$t_{2g}$ and $e_g$ and O-2$p$ states}

\author{Priya Mahadevan$^1$, I. V. Solovyev$^{1,2}$ and K. Terakura$^{3,4}$}
\address{
$^1$JRCAT-Angstrom Technology Partnership,
1-1-4 Higashi, Tsukuba, Ibaraki 305-0046, Japan\\
$^2$Institute of Metal Physics, Russian Academy of Sciences,
Ekaterinburg GSP-170, Russia\\
$^3$JRCAT-NAIR, 1-1-4 Higashi,
Tsukuba, Ibaraki 305-8562, Japan \\
$^4$Institute of Industrial Science, University of Tokyo,
7-22-1 Roppongi, Minato-ku, Tokyo 106-8558, Japan
}
\date{\today}
\maketitle
\begin{abstract}
The low-temperature
spin dynamics of doped manganites have been analyzed within a tight-binding
model, the parameters of which are estimated by mapping the results of
{\it ab initio} density functional calculations onto the model. 
This approach is found to provide a good description of the spin
dynamics of the doped manganites, observed earlier within the {\it ab initio}
calculations.
Our analysis not only provides some insight into the roles of the $e_g$ and the
$t_{2g}$ states but also indicates that the oxygen $p$ states play an important
role in the spin dynamics.  This may cast doubt on the adaptability
of the conventional model Hamiltonian approaches to the analysis of
spin dynamics of doped manganites.
\end{abstract}
\pacs{75.30Ds, 75.10.Lp, 75.30.Et, 75.50.Cc}

There has been resurgence of interest in transition metal oxides
with the perovskite structure
owing to their wide range of electronic and magnetic properties. Among
them, the
hole doped manganites \cite{lamn} have been occupying a special
position: they exhibit dramatic phenomena like colossal
magnetoresistance and are being intensively studied with
prospect for technological applications. LaMnO$_3$, the
parent material of the manganites, is
an antiferromagnetic insulator. Upon sufficient doping ($x \sim$ 0.15)
with divalent ions (such as Sr, Ca), the system is driven metallic.
The holes are allowed to move only if
adjacent spins are parallel, which results in a dramatic increase
in the conductivity when the spins order ferromagnetically, an effect
which can be induced by applying a magnetic field or
by lowering the temperature below the Curie temperature, $T_c$.
Thus, the carrier
mobility is intimately related to the underlying magnetic state of the
system, and there have been considerable efforts in recent times to
identify the interactions that control the
magnetoresistive properties. An approach in this
direction has been to analyze the spin dynamics of the doped
manganites.

Early experiments on La$_{0.7}$Pb$_{0.3}$MnO$_3$ \cite{spinw1} indicated that
the spin-wave dispersion $\omega ({\bf q})$ in the
doped manganites could be interpreted in terms of a conventional
Heisenberg ferromagnet with only the nearest-neighbor exchange coupling.
This behavior is consistent with the double-exchange limit of the
one-band ferromagnetic Kondo lattice model \cite{dex}, implying that
conduction $e_g$ electrons move in a tight-binding band with one orbital
per site and interact with localized $t_{2g}$ spins via the large
intraatomic exchange $J_H$.
However, more recent
experiments \cite{spinw2} on other doped manganites have found
strong deviation of $\omega ({\bf q})$
from the simple cosine-like behavior
expected from a nearest-neighbor Heisenberg model.
Farther neighbor interactions in addition to the nearest neighbor one
had to be taken into account to reproduce
softening of the dispersions for the wave vector ${\bf q}$ approaching
the zone boundary.
The one-band models
could not explain the observed zone boundary softening, even
qualitatively \cite{dex,dex2}.
It was then
suggested that additional degrees of freedom, probably the
lattice degrees of freedom, may play an important role in the
spin dynamics of these materials. Recently \cite{igor}, it was shown
that the softening at the zone boundary has a purely electronic
origin, and could be explained within the
framework of {\it ab initio}  density functional band calculations,
in the local-spin-density approximation (LSDA).

The previous work \cite{igor} also carried out a perturbative analysis of the
exchange interaction strengths within a tight-binding model considering
the double degeneracy of $e_g$ levels
on the Mn site.
It was argued that the degeneracy of $e_g$ orbitals plays important
roles, and simply by taking into account the proper structure of the
kinetic hopping between nearest-neighbor $e_g$ levels one can,
to a large extent, understand the behavior of two strongest interactions,
$J_1$ and $J_4$, in the half-metallic regime.
Here, $J_k$ corresponds to the exchange interaction between the
$k$-th neighbor atoms \cite{convention},
as defined later by Eq.~\ref{eofq}. Furthermore, it was
pointed out that a realistic model including the
oxygen $p$-orbitals and the Mn $t_{2g}$ orbitals could indeed modify the
quantitative aspects of the results. While the fully-filled majority spin
$t_{2g}$ orbitals could contribute an antiferromagnetic superexchange
component to $J_1$, the partially filled minority spin $t_{2g}$ orbitals
could, as suggested by the {\it ab initio}
band structure calculations, contribute a ferromagnetic
double exchange component.
Further $J_2$ was found to
increase quite strongly in the $e_g$-only model, unlike the
weak dependence seen in the
{\it ab initio} results.
The previous
work \cite{igor} suggested that the oxygen $p$ bands could modify
$J_2$ considerably as the Mn 3$d$-O 2p energy separation is comparable
with the exchange splitting of the majority and minority spin orbitals.
In the light of these observations,
we have attempted to make further quantitative analysis
to understand the origin of the observed zone-boundary softening.
This has been done by mapping the results of the {\it ab initio}
band structure
calculations onto a tight-binding model which gives us
flexibility of constructing simpler models and analyzing the
contributions to the observed softening.

The band structure for hypothetical cubic ferromagnetic LaMnO$_3$
with the lattice parameter of 3.934~\AA, calculated
within the linear-muffintin orbital method with the atomic
sphere approximation (LMTO-ASA), was mapped onto a nearest neighbor
tight binding model \cite{tb1} which had been found to give a good description
of the electronic structure of the transition metal oxides of the form
La$M$O$_3$, where $M$=Ti-Ni.
The tight-binding Hamiltonian
consists of the bare energies of
the transition metal $d$ ($\epsilon_d$)
and the oxygen $p$ ($\epsilon_p$) states and hopping interactions
between the orbitals on neighboring atoms. The nearest neighbor
hopping interactions were expressed in terms of the four
Slater Koster parameters, namely $pp\sigma$, $pp\pi$, $pd\sigma$
and $pd\pi$. Note that no direct $d$-$d$ hopping was taken into
account.  While $p$-$d$ covalency effects lift the degeneracy
of the $d$ orbitals, an additional interaction $sd\sigma$
between the transition metal $d$ and oxygen 2$s$ orbitals
was required to lift the degeneracy at the $\Gamma$ point \cite{matth1}.
The energy of the oxygen 2$s$ level was fixed at $-20$~eV.
In order to obtain the magnetic ground state within the single
particle tight binding model, we have introduced an extra
parameter, ($\epsilon_{pol}$) which is the bare energy difference
between the up and down spin $d$ electrons at the same site.
An additional splitting, ($\epsilon_{pol}^{\prime}$) was introduced
between the up and down spin $d$ orbitals of $e_g$ symmetry~\cite{comm0}. The
parameters entering the tight-binding Hamiltonian were
determined by the least square fitting of the energies
obtained from tight-binding calculations
at several $k$-points to those
obtained from the LMTO calculations. It should be noted that the
deep lying oxygen 2$s$ bands were not involved in the fitting.
The extracted parameters are $sd\sigma=-1.57$~eV, $pp\sigma$=~~0.91~eV,
$pp\pi=-0.23$~eV, $pd\sigma=-2.02$~eV, $pd\pi$=~~1.0~eV,
$\epsilon_d - \epsilon_p$=0.48~eV,
$\epsilon_{pol}$=3.2~eV, $\epsilon_{pol}^{\prime}$=0.3~eV being consistent
with the earlier estimate for the system \cite{tb1}.

The frozen spin spiral approximation \cite{sandratskii}, where
the orientation of the magnetic moment at each atomic site is
spirally modulated by the wave vector ${\bf q}$, was used
to calculate the exchange interaction $J_{\bf q}$ defined by
\begin{equation}
J_{\bf q} = \sum_{k} J_k \exp [i{\bf q} \cdot {\bf R}_k ] \\
\end{equation}
where ${\bf R}_i$ is the position vector of $i$-th Mn atom.
$J_k$ is the $k$-th neighbor exchange interaction
appearing in the Heisenberg Hamiltonian given by
\begin{equation}
E[\{{\bf e}_i \}] = -\frac{1}{2} \sum_{ik} J_k {\bf e}_i \cdot
{\bf e}_{i+k}
\label{eofq}
\end{equation}
with {\bf e}$_i$ denoting the
direction of the magnetic moment at the
site $i$. By using the local force
theorem \cite{lft}, the changes in the single particle energy could be
related to the exchange interaction by  mapping
onto the Heisenberg model as defined above.
A rigid band picture was adopted to simulate the doping effects~\cite{comm2}.
Simplified models were constructed to elucidate the mechanism of
zone-boundary softening of the spin wave,
and comparison was made with the results from the LMTO
calculations whenever possible to ensure that the present result is not an
artifact of a particular parameter set.

In Fig.~1a we show the LMTO results for the spin
dispersion $\omega ({\bf q}; x)$
along the symmetry directions $\Gamma$X, XM and MR calculated
for several doping values $x$. In the small ${\bf q}$ region, the 
spin excitations
have a weak dependence on doping as is evident from the
result along the $\Gamma$X direction. However, sufficiently away from the
$\Gamma$ point, the results become a strong function of the concentration $x$.
Considering the result for $x$=0.4 along the $\Gamma$X
direction, we see that the spin dispersion is almost flat
from midway to the zone boundary. The experimental result for
Pr$_{0.63}$Sr$_{0.37}$MnO$_3$ \cite{spinw2}
along $\Gamma$X
shows very similar behavior.
The results of the tight-binding
model, calculated by using the parameters extracted by fitting the
{\it ab initio} band structure are shown in Fig.~1b.
This model calculation is called model A in order to distinguish from other
models discussed later.
Model A is seen to provide a good description of the energetics
of the spin dynamics observed within the {\it ab initio}
approach. The results in Fig.~1 suggest that the difference
in energy between the ferromagnetic ground state and the various
commensurate antiferromagnetic (AF) spin configurations such as
those defined at the X (A-type AF), M (C-type AF)
and R (G-type AF) points decreases with doping.

The Fourier transform of $J_{\bf q}$ gives us the real space
exchange integrals $J_i$ as shown in Fig.~2.
Dominant interactions are all confined within the linear
-Mn-O-Mn-\ldots chains parallel to $<001>$ as was pointed out
already~\cite{igor}.  They are $J_i$ with $i$=1, 4 and 8.
$J_2$ is the interaction for the pairs along $<110>$ and
takes relatively small values partly because of the
cancelation between the contribution from the Mn $d$ bands and that
from O $p$ states.
These results are consistent with the analysis of the experimental
results \cite{spinw2} which required finite $J_4$ and $J_8$ to be
included in the Heisenberg Hamiltonian in order to reproduce the experimentally
observed spin wave dispersions.
In order to understand
the behavior of the spin-wave
dispersion $\omega ({\bf q};x)$ in terms of $J_i$,
the following expressions for ${\bf q} \parallel x$ will be useful.
\begin{eqnarray}
\hbar \omega (q_x)&\simeq& 2 [(J_1 + 4 J_2) \sin^2 \frac{1}{2} q_x a +  J_4
\sin^2
q_x a +
 J_8 \sin^2 \frac{3}{2} q_x a ],
\label{Eq:spinw}
\end{eqnarray}
where $a$ is the lattice constant of the cubic unit cell.
For $q_x a \ll 1$  the above expression reduces to
\begin{eqnarray}
\hbar \omega (q_x)&\simeq& \frac{1}{2} [J_1 + 4 J_2 + 4 J_4 + 9 J_8] (q_x a)^2
\end{eqnarray}
which helps us understand the  weak dependence of the low energy
excitations on the concentration
$x$.
The large prefactors for $J_4$ and $J_8$ indicate that modest
changes in $J_4$ and $J_8$ are sufficient to offset the large
changes in $J_1$ found within our model. At the X point,
Eq.~\ref{Eq:spinw} reduces to
\begin{eqnarray}
\hbar \omega (q_x=\frac{\pi}{a})&\simeq& 2 J_1 + 8 J_2 + 2 J_8.
\end{eqnarray}
As the dependence of $J_2$ on $x$ is weak, the changes at the zone boundary
are driven by $J_1$ and $J_8$. Unlike in the low $q$ regime, the prefactors
of $J_1$ and $J_8$ are equal in this case. Since
the decrease in $J_1$ is much larger
than the increase in $J_8$, the energy at the X point decreases as the
hole concentration is increased. Another useful information about the
flattening of the dispersion beyond half way to the zone boundary is given by
comparing
Eq.~5 with
\begin{eqnarray}
\hbar \omega (q_x=\frac{\pi}{2a})&\simeq& J_1 + 4 J_2 + 2 J_4 +  J_8.
\end{eqnarray}
The energies at $q_x=\frac{\pi}{a}$ and $q_x=\frac{\pi}{2a}$ are
comparable when $J_1$ $\sim$ 2$J_4$.

We constructed simpler models to make quantitative estimates for the
contribution from the $t_{2g}$ electrons and that from the $e_g$
electrons. Model B(C) includes $e_g$ ($t_{2g}$) orbitals on the Mn atoms
and all $p$ orbitals on the oxygens. As the density of states (DOS) obtained
within model A suggests partial occupancy of the minority spin $t_{2g}$ bands
with $\sim$ 0.175 electrons even for the undoped case,
the Fermi energy ($E_{\rm F}$) of model C in the
undoped case was adjusted so that the minority spin $t_{2g}$ bands had
0.175 electrons. As a consequence, the number of holes in the majority
spin $e_g$ states of model B in the undoped case should be larger by
0.175 than the case when the minority spin $t_{2g}$ states are not occupied.
The $d$ partial DOS for models B and C along with the result for model A
which considers all $d$ orbitals on the Mn atom are shown in Fig.~3.
The reduced models (B and C) are found to give
a good description of the respective $d$ partial DOS of $e_g$ and $t_{2g}$
symmetry within model A. Further justification for the treatment of the
contributions from $e_g$ and $t_{2g}$ states separately is given by the
fact that the dominant  contributions, ($J_i$, $i$=1,4,8), are all
for the pairs along $<100>$ for which there is no mixing of the two states
in the exchange coupling.
The spin wave dispersion was
calculated in the reduced models separately.
The dispersion along the
$\Gamma$X direction for model B is shown in the panel b
(inset of Fig.~3a).  $y$ denotes the number of doped $e_g$ holes with
reference to the half filled majority spin
$e_g$ band.  (Equivalently, $1-y$ is the
number of electrons in the $e_g$  band.)
$x$ in the parentheses indicates the hole concentration in model A being
equivalent to doping of divalent atoms.
By considering the above
situation, $y$=0.175 corresponds to the case of undoped LaMnO$_3$, i.e., $x=0$.
As the number of holes increases from $y=0.175$, the spin-wave
energy at the X point steadily decreases.
In model C, on the other hand, the dominant contribution from the
$t_{2g}$ states to the exchange coupling is antiferromagnetic
superexchange.  The negative spin-wave energy for all $q$ (Fig.3d) is
consistent with this expectation.  However, small occupation of the
minority spin $t_{2g}$ states produces a ferromagnetic double exchange
contribution.  $z$ in Fig.3d denotes the number of electrons in the
minority spin $t_{2g}$ states.  Clearly, doping of divalent elements
reduces $z$ so that the double exchange contribution diminishes
rapidly as is clearly seen in the $z$ dependence of the spin-wave
dispersion in Fig.3d. Here again the corresponding value of $x$ in model A
are indicated in brackets.
The contribution from the $e_g$ states (Fig.3b) and that from the $t_{2g}$
states (Fig.3d) for the common $x$ value are added and the resultant
spin wave dispersion shown in Fig.3e agrees very well with the one
in Fig.1b (model A).  This analysis suggests that the main source of the
zone-boundary softening of the spin wave dispersion by doping of
divalent atoms for $x<0.3$ is
the reduction in the ferromagnetic double exchange of the $t_{2g}$
electrons.  On the other hand, in the doping range of $x>0.3$, the
$t_{2g}$ states may simply act as a source of antiferromagnetic superexchange
and further softening and flattening of the spin wave dispersion comes from
the $e_g$ states.

The doped holes within our model have considerable oxygen $p$
character, and the earlier results \cite{igor} suggested that the
itinerant oxygen band could modify the various exchange interaction
strengths.
It was pointed out that the role of oxygen $p$ states
in the superexchange interaction is not only to mediate the $d-d$ transfer
but also to make a direct additional contribution \cite{oguchi,zaanen}.
However in this treatment
the banding effect of oxygen $p$ states was neglected.  This assumption
will not be justified for quantitative arguments
if the $p$ band width is comparable to the
$p-d$ energy separation, which is the case in our systems.
In order to obtain information about the role of the
oxygen $p$ band,
we made  further simplification in the model B
that the hopping between oxygen atoms was neglected, {\it i.e.},
$pp\sigma$=$pp\pi$=0 (model D). As the neglect of the hopping between
the oxygen atoms could reduce the $e_g$ bandwidth, the spin wave
dispersions were calculated for two
values of $pd\sigma$ - the value ($-2.02$~eV) estimated
already by the fitting (results
shown in Fig.~4a) and an increased value of $-2.25$~eV
(results shown in Fig.~4b).
The dispersions shown in Figs.~4a and 4b are qualitatively similar
to each other. In both cases, the spin wave energy at the X point
increases with doping being in contradiction to the behavior observed
in Fig.~1 and Fig.~3b.
Since this model does not take into account the antiferromagnetic
superexchange contributions coming from the $t_{2g}$ degrees of freedom
and affecting primarily the nearest-neighbor magnetic interactions,
the ferromagnetic coupling $J_1$ remains to be the strongest interaction
in the system. In such a situation, the form of the spin-wave dispersion
is close to the cosine-like.

To analyze further the role played by
the inter-oxygen hopping,
the exchange interactions \{$J_i$\} were obtained for the cases
corresponding to Fig.~3b and Fig.~4. The results are shown in Fig.~5.
As is expected, the behavior of $J_1$ is not affected so much by the $p-p$
hopping.
By comparing the results of model A in Fig.2  with those
of model B in Fig.5, we see that
the doping dependence of $J_4$ and $J_8$ comes primarily from
the $e_g$ electrons.  Neglect of the $p-p$ hopping (model D) strongly
suppresses
$J_4$ and $J_8$ for $x$ $\ge$ 0.2, and zone boundary softening of the
spin-wave dispersion become less pronounced.
On the other hand,  increase of $J_2$ with hole doping is enhanced by
neglecting the $p-p$ hopping.  The energy at X point
increases in Fig. 4 (model D) with hole doping, because the variation
in $J_1$ in this case is not enough to offset the sharp increase
in $J_2$.

In summary, we have analyzed the low-temperature
spin dynamics of the doped
manganites with tight-binding models. Our results
provide some insight into the roles of the $t_{2g}$ and the $e_g$
states and also suggest that the channel of hopping between the
oxygen atoms strongly modifies the exchange interactions.
Thus for the correct
quantitative and sometimes even qualitative
description, the simplifications made by model Hamiltonian approaches
which consider only the $e_g$ orbitals are questionable.

We thank Prof.~D.D.~Sarma for useful discussions. Part of the programmes
used here were developed in Prof.~Sarma's group.
The present work is partly supported by NEDO.

\pagebreak
\section{figure captions}

Fig.1  The spin wave dispersions obtained by (a) LMTO calculations and
(b) the tight binding approach (model A)
along the symmetry directions $\Gamma$X,
XM and MR shown as a function of doping.

Fig. 2 The doping dependence of the exchange couplings $J_1$,
$J_2$,
$J_4$ and $J_8$ between atoms at ($a$ 0 0), ($a$ $a$ 0), (2$a$ 0 0) and
(3$a$ 0 0),
where $a$ is the lattice parameter.

Fig. 3 The (a) minority-spin and (c) majority-spin $d$ partial density of
states
within models A, B and C. The spin wave dispersions along
$\Gamma$X as a function of doping within models  (b) B
and (d) C  are shown along with (e) the combined contributions
of models B and C. $y$ refers to the hole concentration in the majority-spin
$e_g$ band with reference to its half-filled case.  $z$ is the electron
concentration in the minority-spin $t_{2g}$  band.  $x$ is the net
concentration
of the doped holes and is given by $x=y-z$.

Fig. 4 The dependence of the
spin wave energies on the $e_g$ hole doping $y$ along $\Gamma$X
within model D.  The hopping between oxygen atoms and the
$t_{2g}$ orbitals on the Mn atom have
been left out of the model.
(a) $pd\sigma$=-2.02~eV
and (b) $pd\sigma$=-2.25~eV.

Fig. 5 The variation of the exchange couplings $J_1$, $J_2$,
$J_4$ and $J_8$ with the $e_g$ hole doping  $y$.
Open circles are for the case (model B) including the
hopping between oxygen atoms and $pd\sigma$=-2.02~eV.
Open and filled squares are for the cases (model D) without the hopping
between oxygen atoms: $pd\sigma$=-2.02~eV (open squares)
and $pd\sigma$=-2.25~eV (filled squares). The $t_{2g}$ orbitals on the
Mn atom have been left out of the basis set.


\begin{references}
\bibitem{convention}
$J_{\it 1}$, $J_{\it 2}$, $J_{\it 3}$,
$J_{\it 4}$ and $J_{\it 8}$
are the interactions in the cubic lattice
corresponding to the interatomic vectors
$[0,0,a]$, $[a,a,0]$, $[a,a,a]$,
$[0,0,2a]$ and $[0,0,3a]$, where $a$ is the cubic lattice
constant.
\bibitem{comm0}
The parameter $\epsilon_{pol}^{\prime}$ is anyway small and
turns out not to be important in the following calculations.
\bibitem{comm2} i.e., we start with the undoped cubic ferromagnetic
LaMnO$_3$, whose electronic structure has been obtained self-consistently,
and shift the position of the Fermi level in order to adjust the required
hole-concentration $x$, without an additional self-consistency. Note,
that this is different from the procedure used in Ref.\cite{igor},
where the electronic structure of the hypothetical virtual-crystal
alloy was calculated self-consistently in the framework of LSDA for
each concentration $x$.
\end{references}
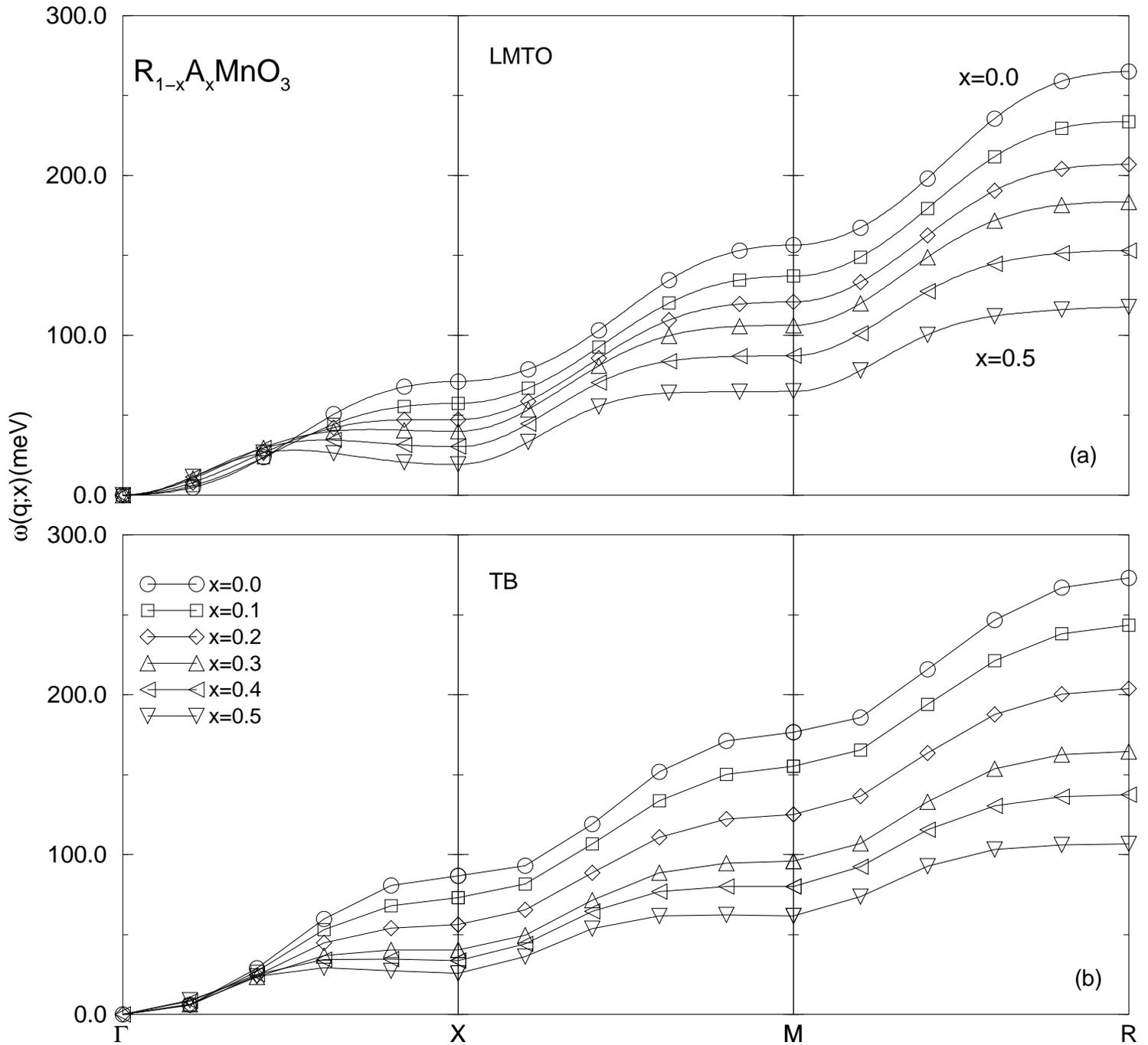
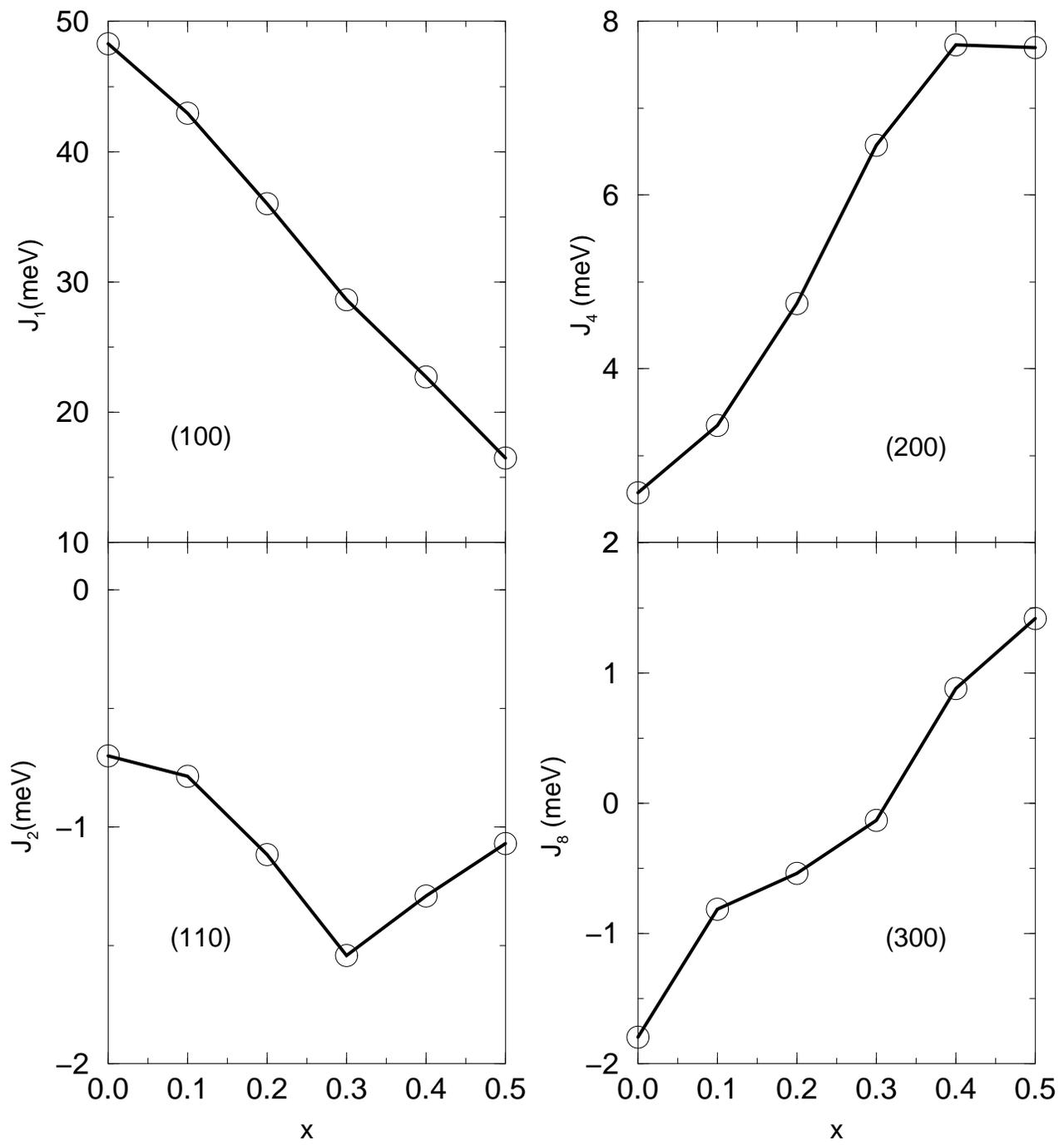
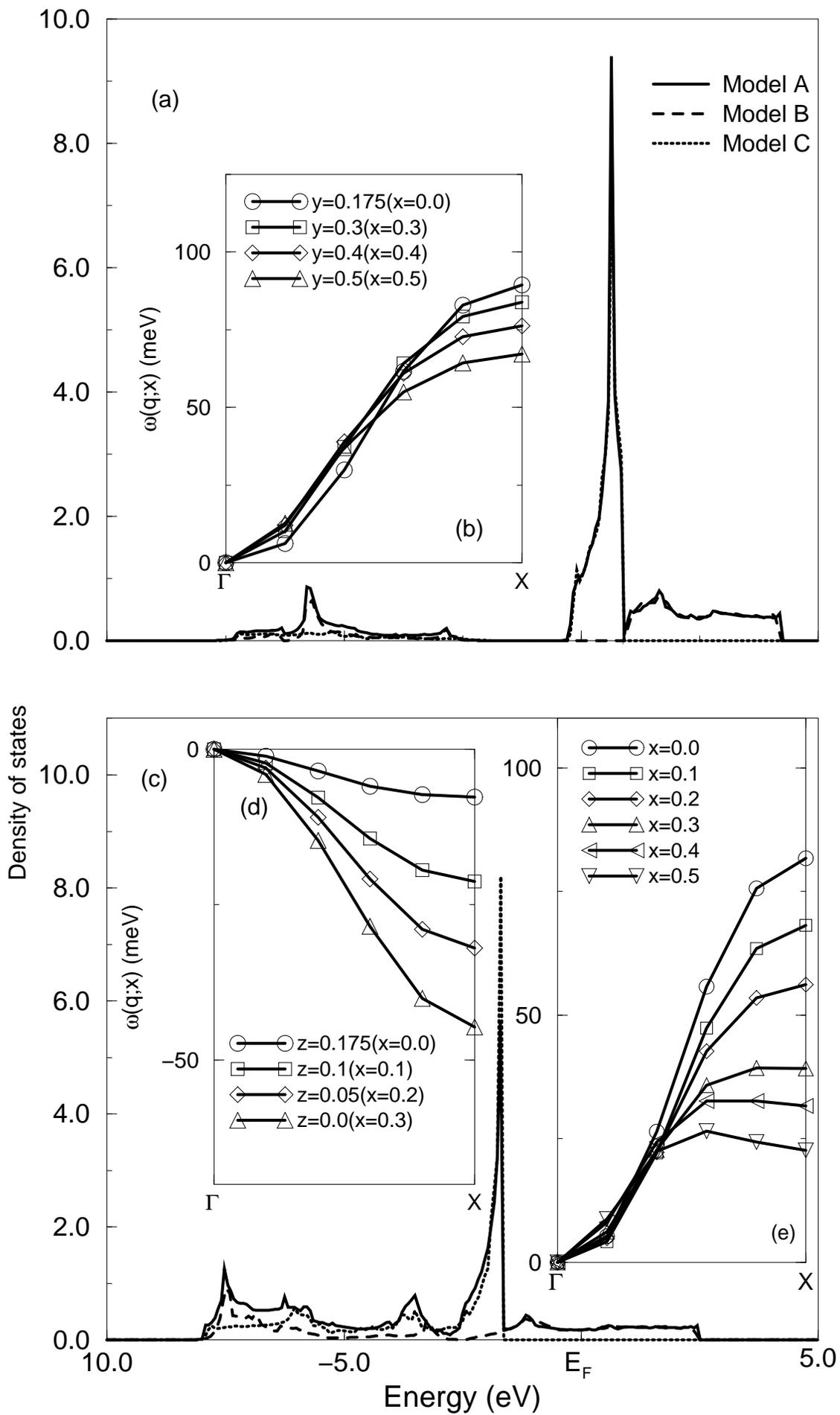
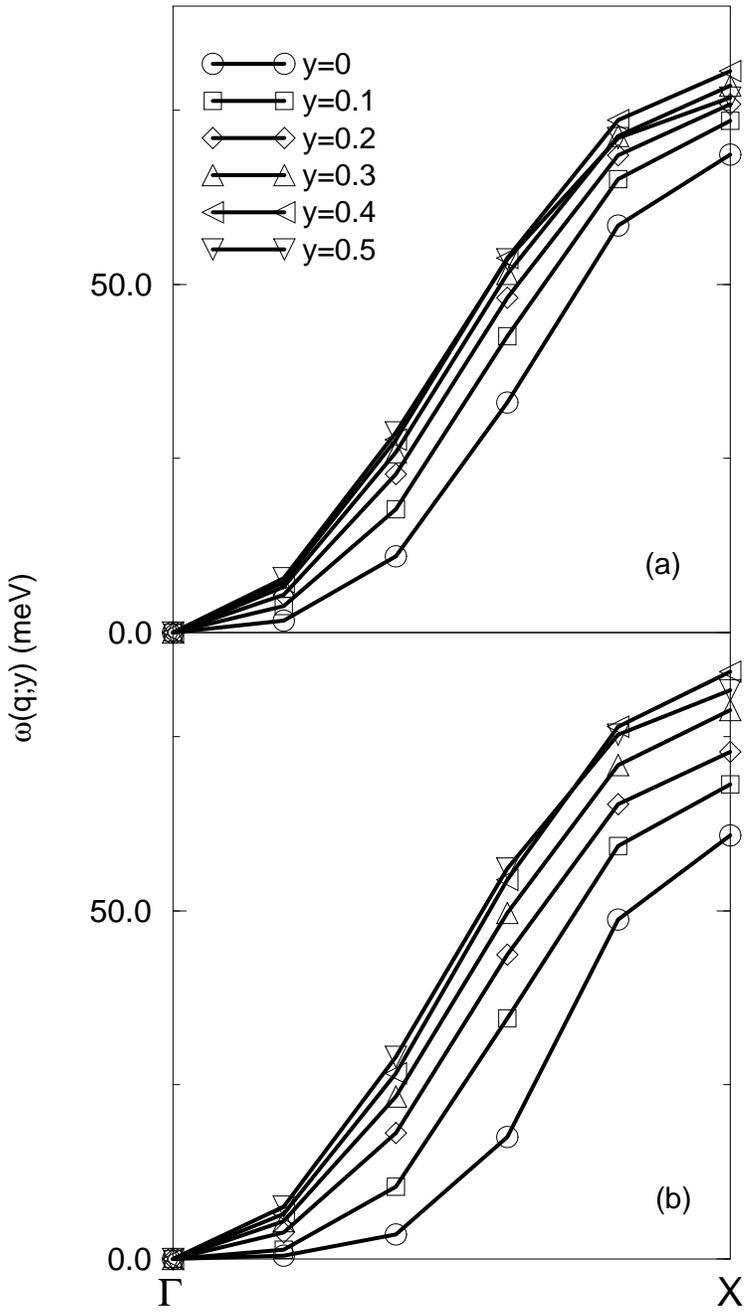
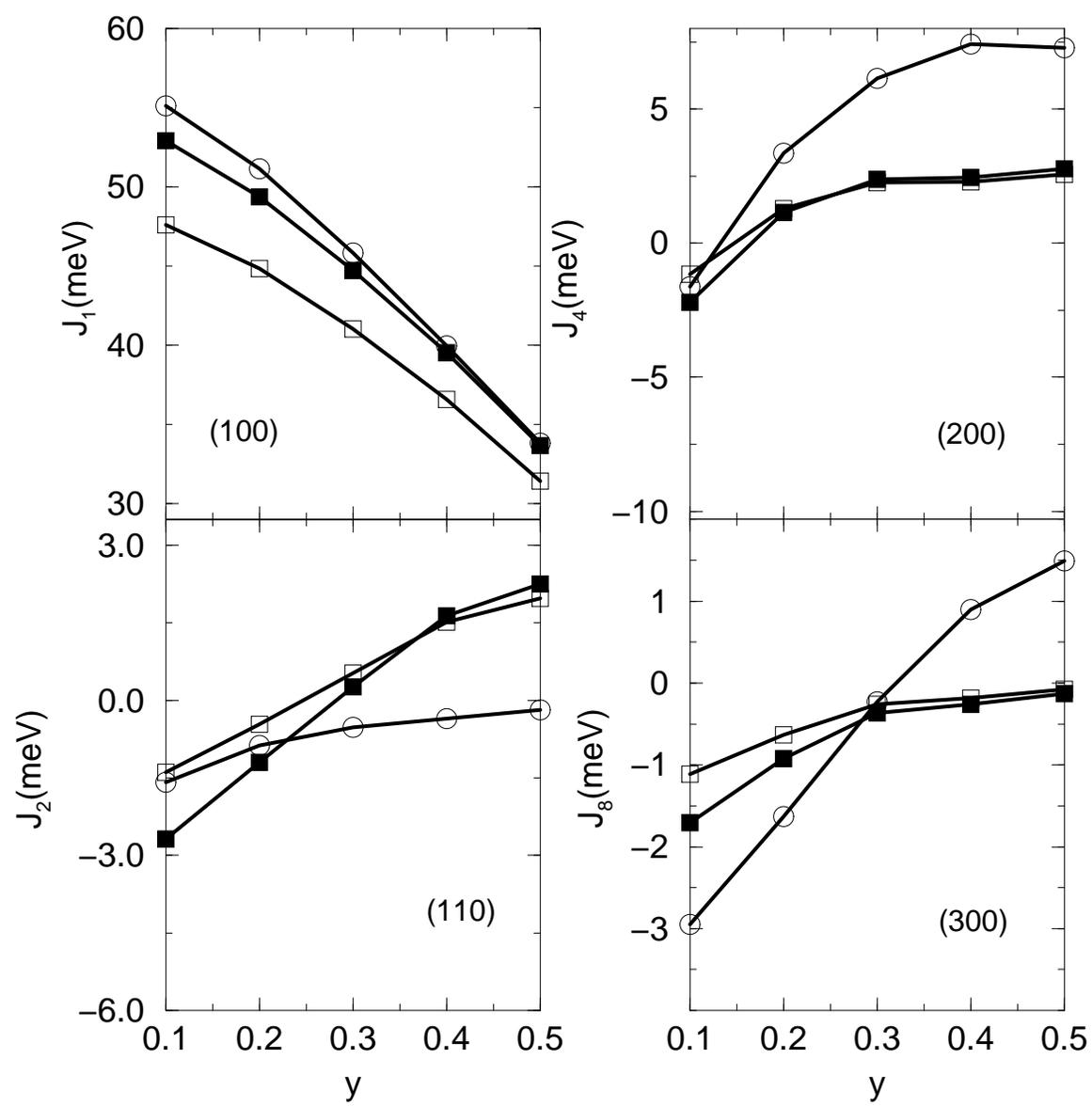
\end{document}